# Quantum Science and Technologies in K-12: Supporting Teachers to Integrate Quantum in STEM Classrooms

Nancy Holincheck[1] *, Jessica L Rosenberg[2], Xiaolu Zhang[1], Tiffany N Butler[1], Michele Colandene[1] and Benjamin W Dreyfus[2,3]

[1] College of Education and Human Development, George Mason University, Fairfax, VA 22030, USA; xzhang22@gmu.edu (X.Z.), tbutle5@gmu.edu (T.N.B.), mcolande@gmu.edu (M.C.);

[2] Department of Physics and Astronomy, George Mason University, Fairfax, VA 22030, USA; jrosenb4@gmu.edu (J.L.R.), bdreyfu2@gmu.edu (B.W.D.)

[3] STEM Accelerator Program, George Mason University, Fairfax, VA 22030, USA; bdreyfu2@gmu.edu (B.W.D.)

\* Correspondence: nholinch@gmu.edu;

**Abstract:** Quantum science and computing represent a vital intersection between science and technology, gaining increasing importance in modern society. There is a pressing need to incorporate these concepts into the K-12 curriculum, equipping new generations with the tools to navigate and thrive in an evolving technological landscape. This study explores the professional learning of K-12 teachers ($n$ = 49) related to quantum concepts and pedagogy. We used open-ended surveys, field notes, workshop artifacts, and interviews to examine teachers' perceptions of quantum and how they made connections between quantum and their curriculum. Our data reveal that most teachers were excited and interested in teaching quantum but were aware of potential barriers and concerns that might get in the way of teaching quantum. We found that teachers readily identified connections to math and science in their curriculum, but only a few made connections to computing. Enthusiasm for teaching quantum concepts was found in both elementary and secondary educators, suggesting a widespread recognition of its importance in preparing students for a future where quantum technology is a fundamental aspect of their lives and careers.

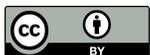

## 1. Introduction

Quantum science, computing, and its applications are becoming increasingly important in modern society, yet quantum concepts are largely unaddressed in the standard K-12 curriculum. There is little research on teaching K-12 quantum, as quantum has historically been accessible only to advanced physics students. However, in the near future, quantum technologies will impact individuals and industries across and beyond the STEM fields, signaling a paradigm shift that extends from redefining computing and information processing to altering our philosophical understanding of reality. Quantum cryptography, sensors, and computing will radically change a broad range of STEM fields, including medicine, national security, materials science, data science, and additional quantum-adjacent fields [1,2]. Quantum knowledge and technologies have the potential to become as commonplace as computer knowledge has become for us over the past decades.

One contributing factor to the perceived complexity of quantum concepts is their "counterintuitive" nature, which often clashes with intuitions developed through years of traditional education [3]. Although quantum has applications across and beyond the STEM disciplines, quantum science is most often associated with quantum mechanics, a branch of modern physics that pertains to the structure of atoms and behavior of subatomic particles [4]. Quantum mechanics explains empirical observations that conflict with classical physics [5]. Quantum physics describes the behavior of subatomic particles in terms of probabilities rather than deterministic equations [6]. It describes seemingly illogical properties of light and matter, including wave–particle duality and superposition of states. Heisenberg [7] noted that the new ideas would "cause a revolution in thinking and therefore concern a wide range of people" (p. 9).

Over the past century, quantum has indeed changed our world, as quantum science has been credited for the invention and application of transistors, lasers, and computers [8]. More recent advances include the ongoing development of quantum sensors, quantum computing, and quantum communications [9]. In light of these advancements, K-12 schools must adapt curricula to familiarize the younger generations with quantum concepts from an early age. In the context of the growing importance of quantum in our understanding of the world, it becomes essential to identify effective pedagogical methods for integrating these ideas into K-12 education. The challenge lies in making quantum concepts accessible and engaging to younger students. This task, while pioneering, is crucial in preparing them for a future increasingly influenced by quantum science. Introducing quantum concepts beginning at an early age can also spark curiosity in STEM among a diverse student population in U.S. classrooms, which may help to address the racial, ethnic, and gender imbalances evident within the U.S. STEM workforce [10–12]. There is great potential for engaging learners through and with emerging technologies like quantum; but, before we can be ready to teach quantum to K-12 students, we must first understand how to prepare K-12 teachers to do this. This study explores teachers' professional development experiences learning quantum content and pedagogy.

## 2. Literature Review

The National Quantum Initiative, signed into law in December 2018, mandated the creation of new research and educational programs to prioritize the development of a new high-tech workforce [13,14]. Quantum workforce development remains a top bipartisan priority with a focus on elementary and secondary education [15,16]. Despite this push, few curricular resources are available for K-12 quantum education, and little is known about how to prepare K-12 teachers to teach quantum content. Quantum education research has traditionally focused on undergraduate student learning in physics courses [17–21]. At the K-12 level, research has predominantly focused on high school physics, with some additional research on computer science applications of quantum [19,22].

Although quantum originated within the physical sciences, it is now a transdisciplinary STEM topic, with researchers across mathematics, computing, engineering, life sciences, and physical sciences researching theory and applications [23,24]. Quantum concepts and applications span the STEM disciplines and beyond [23]. Recent work by the National Q-12 Education Partnership [15] led to the development of frameworks for integrating quantum concepts into specific disciplines within K-12 education. The Q-12 Partnership has released frameworks for middle school STEM and high school computer science, physics, chemistry, and mathematics. These frameworks are intended to support curriculum developers and teachers by providing guidance about how discipline-specific learning goals could be met through teaching quantum concepts. These frameworks are all structured around nine key concepts: (a) quantum information science, (b) quantum state, (c) quantum measurement, (d) qubit, (e) entanglement, (f) coherence, (g) quantum computers, (h) quantum communication, and (i) quantum sensors.

K-12 STEM concepts that can be taught within or through quantum concepts include mathematical probability and probabilistic thinking, vectors, matrices, atomic structure, atomic energy levels, conservation of energy, waves and optics, binary digits (bits), computer storage, and electricity. As K-12 students learn quantum concepts, they can engage in science and engineering practices [25] and mathematical practices [26], especially asking questions, problem-solving, abstract reasoning, developing and using models, and constructing explanations. STEM learning experiences centered on problem-solving, modeling, and simulation [27] allow students to engage in sense-making around authentic, real-world problems and help them connect new STEM learning to their prior experiences. The complex, counterintuitive nature of quantum concepts makes it particularly important that students develop mental models that support their thinking about quantum [28].

*2.1. Integrating Quantum into K-12 STEM*

There are multiple arguments for teaching quantum in K-12 schools, including preparing students to work in current and future quantum jobs [29-31], developing quantum literacy in general [24], as well as educating "the student as a person rather than as a potential scientist" [32] (p. 2). Our research team recognizes the need to prepare the future quantum workforce, and we are committed to building educational programs to prepare students for quantum and quantum-adjacent careers. Yet, our interest in K-12 quantum is also focused more broadly on developing teacher and student quantum literacy, defined as understanding quantum and its applications across "diverse areas of society" [24] (p. 566).

Recent research offers insight into what quantum learning can look like in K-12. Student learning of quantum concepts and applications can occur outside of school through extracurricular clubs, like that described by Silberman [33], who used online resources and field trips to university labs in his work with high school students. Within K-12 classrooms, researchers with the Quantum for All project have conducted workshops for high school students and physics teachers [34]. Some of the teachers in this project felt they would need multiple rounds of professional development to be ready to integrate quantum into their curriculum, pointing to the complexity of the concepts.

One concern noted in the literature is that mathematics can hinder students from accessing quantum [35]. In recognition of this barrier, Dündar-Coecke and colleagues [35] have recently promoted "Quantum Picturalism" as a novel approach to teaching quantum concepts.

This method rigorously teaches quantum concepts but reduces the complexity of the mathematics to support student understanding.

A recent review of the K-12 quantum education literature [18] identified recommendations for K-12 quantum computing education. These included (a) relating quantum to everyday events, (b) employing active learning strategies, (c) focusing less on mathematics formulas and more on concepts, and (d) aligning quantum content with existing standards. These are echoed in Hasanovic's [36] recommendations for quantum instruction to use a visual, hands-on approach that incorporates analogies to make the content more accessible. Choudhary and colleagues [37] suggested that some quantum concepts could be taught at the middle school level, as did Farris et al. [38], who studied the development of middle school curricular activities related to using quantum methods for drug discovery. Franklin et al. [39] described an NSF-funded project in which they identified initial learning trajectories for young learners in quantum education. They also developed curricular resources for examining the concept of quantum reversibility with third-grade students [40].

In one study set in the United Kingdom, researchers found that using a computer puzzle game to teach students aged 11 to 18 to construct algorithms for use with quantum computers resulted in greater interest in quantum physics and curiosity to learn the mathematics behind quantum computing [41]. Seskir and colleagues [42] reviewed a range of quantum games and other interactive tools used to teach quantum concepts across STEM disciplines, noting that games have the potential to reach students beyond traditional educational settings.

Foti and colleagues [3] also promoted gamification in their work to bring quantum concepts into K-12 education in Finland. Their method focuses on play, discovery, and learning as key strategies for students of all ages to grasp quantum concepts. While the tools for measuring the effectiveness of this approach are still evolving, this study underscores the revolutionary potential of using creative pedagogical methods to teach quantum mechanics. This team's work highlights a crucial distinction between classical and quantum physics education. In classical physics, principles can be demonstrated through everyday phenomena, making them more tangible and relatable [3,43]. However, teaching quantum concepts necessitates a unique approach to teaching and comprehension due to their less-observable and less-relatable nature. This pivot in educational strategy is essential for effectively conveying the abstract and complex nature of quantum mechanics to students.

*2.2. K-12 Students' Cognitive Readiness for Quantum*

Within science education, there is a history of pointing to Piaget's theory of cognitive development as evidence that abstraction is too challenging for young learners, as they are only ready for concrete experiences [44]. However, education researchers have established that students in early elementary grades can engage in abstract thinking and that this ability continues to develop as students age [45]. A recent review of U.S. educational standards illustrated that abstract thinking during science learning is expected of students from a young age [46]. Similarly, neuroscience research has demonstrated that students as young as preschool age are able to engage in abstract thinking and that their ability to do so improves as they develop a more robust working memory, improved information processing, and additional disciplinary knowledge [47]. The abstract nature of quantum concepts necessitates the use of analogies, as illustrated by several studies focused on specific analogies used in quantum (e.g., see [48,49]).

Engaging teachers and students with quantum can harness the potential for quantum to engage each and every student in STEM. It also provides students with exposure to quantum concepts and vocabulary so they will be better prepared to engage with these ideas in the future [36]. Learners find quantum exciting and new. We can leverage the novelty of quantum to provide opportunity and access to diverse students who are under-represented in the current STEM workforce to build a new pipeline to a future diverse quantum workforce [30]. Quantum concepts provide space to build students' thinking about the nature of science, challenging our unreliable intuition regarding math and science [23] and expanding perspectives on what it means to know and do science. To bring this new topic that has, historically, only been taught to upper-level undergraduate and graduate students to K-12 schools will require a new model for teacher engagement and preparation.

## 3. Research Methods

In this exploratory study, we used a critical realist approach [50], utilizing qualitative methods [51] to explore K-12 teachers' perceptions of quantum, how they developed an understanding of quantum concepts while engaged in quantum professional learning, and how they connected quantum content and their current grade-level curriculum. Theorists position the critical realist methodological perspective between positivism and constructivism, combining a positivistic realist ontology with epistemological constructivism. Critical realism assumes that there is a real world independent of our perceptions but that our understanding of the world—including our experiences and beliefs—is not objective [50]. A critical realist approach to research allows researchers to explain events and outcomes within the contexts in which data are collected and analyzed. In this study, qualitative methods allowed us to collect a rich data set and to preserve the context of the data.

The following research questions guided this study.

1. What are K-12 teachers' perceptions of incorporating quantum in their teaching?
2. How do K-12 teachers make connections between their K12 curriculum and quantum concepts?

### 3.1. Context of the Study

This study is situated within a multi-faceted project that seeks to address K-20 quantum workforce development. The overarching goal of the larger grant-funded project is to pilot efforts designed to inspire the next generation of students to pursue quantum in our region. The study described in this paper aimed to investigate teacher learning of quantum concepts and pedagogy. During the 2022–23 academic year, we led six teacher professional learning workshops on quantum concepts and quantum teaching. Specific quantum concepts addressed in our professional learning included the probabilistic nature of quantum, quantum measurement, quantum superposition, and quantum entanglement.

We worked with school district personnel to recruit K-12 teachers from around the mid-Atlantic region of the U.S. to attend our quantum workshops. We held five in-person and one online workshop. The in-person workshops were held on Saturdays for five hours each; participants attended one of these workshops. In these workshops, we first led teachers in interactive professional learning around quantum phenomena, incorporating visuals, hands-on materials, and a quantum-related board game created by our team for the workshops. Teachers were then invited to explore curricular resources, which included discrete activities on singular concepts, online games, and more complex exercises. Our online workshop was a two-hour workshop held over Zoom and included a presentation about quantum concepts, followed by small group work with online resources related to quantum. Before the workshop, participants were sorted into small groups by the grade level they taught (elementary, middle, high). Secondary teachers were also sorted by discipline (physical sciences, earth and biological sciences, and computing). Small group work was facilitated by faculty and students working on the project.

The research team included physics professors, STEM education faculty, doctoral students in science education research, and undergraduate STEM majors. Early in the project, we made assumptions about what we would find. We all initially believed that the quantum professional learning workshops would be of greater interest to high school teachers than elementary teachers and that they would find more connections to their disciplines than elementary and middle school teachers. As discussed below, this was not the case.

*3.2. Participants*

A total of 70 K-12 teachers participated in our quantum professional learning during the 2022–23 academic year, 49 of whom consented to participate in our research study. Of those agreeing to participate in the study, 24 participated in an in-person workshop, and 25 participated in an online workshop. Participants included 17 elementary teachers, nine middle, and 22 high school teachers. The group was majority white (n = 26) but also included six Hispanic teachers, three Black teachers, three Asian teachers, five teachers who identified with multiple races, and six who chose not to share their racial/ethnic identity. The group comprised 34 women, 14 men, and one non-binary teacher.

We conducted follow-up interviews with seven of these teachers. This group included four White women who taught elementary, one White woman who taught high school physics, one Black woman who taught high school math and computer science, and one White man who taught high school physics.

*3.3. Data Collection*

Data collected in this study included researchers' field notes collected during and after the professional learning workshops, collaborative artifacts from the workshops (e.g., posters created when sharing group work and discussions), an exit survey that included open-ended and demographic questions, and follow-up interviews with a sub-group of teachers who agreed to participate in an interview.

Researcher field notes were captured by the first author intermittently during the workshops and then transcribed and expanded upon within a running researcher memo following each professional development session. The exit survey was constructed by the research team, following the recommendations of Dillman et al. [52] for survey construction. Due to the short amount of time we had with teachers, we elected to use only an exit survey. The majority of questions on the survey were open-ended, as we were interested in meaningful qualitative data to use in this study. We included questions designed to leave space for both positive perceptions (e.g., What is exciting about teaching quantum in your classroom?) and negative perceptions (e.g., What concerns you about teaching quantum in your classroom?). To try to understand teachers' understanding and expectations before and after attending the workshop, we also asked a series of questions that incorporated the "I used to think…, but now I think…., and so I will…" thinking routine from Project Zero [53]. This thinking routine was chosen to elicit reflections from teachers about their learning and positioned them to consider and imagine future actions. Additional open-ended questions asked teachers to make connections to the content they teach and asked about additional supports our research team could provide to support them in integrating any of the activities they had experienced during the professional development. Teachers were also asked if we could contact them for a follow-up interview.

We interviewed a subset of teachers (n = 7) using a semi-structured interview protocol that included questions written by the research team. Interviews were conducted via Zoom teleconference, recorded using the Zoom recording tool, and transcribed using Descript (version 74) automatic transcription software. A member of the research team reviewed each audio recording as they carefully checked each transcript and then corrected any errors in the automatic transcription. Interview questions focused on the teachers' experiences during the quantum professional development, their interest in quantum and quantum education, their sense of readiness to teach quantum in their classroom, and how they envisioned using the concepts they had learned and resources provided to them in the quantum professional development.

*3.4. Data Analysis*

We employed thematic qualitative analysis techniques [54], including constant comparative methods [55,56]. The data were iteratively reviewed by two of the authors, generating initial codes of the data. Both descriptive coding and in vivo coding methods were used in the initial cycle of analysis [57]. All data were coded by at least two members of the research team. Each of us independently reviewed the data in advance of our meetings and then engaged in collaborative and reflexive coding during multiple research meetings, focused on developing a nuanced reading of the data [58]. In our consolidation of codes and identification of themes, we followed a critical realist approach to thematic analysis with the goal of developing experiential themes, inferential themes, and dispositional themes [59].

**4. Findings**

Our data reveal that most teachers were excited about and interested in teaching quantum but were aware of potential barriers and concerns that might get in the way of teaching quantum. We found that teachers readily identified connections to math and science in their curriculum, but only some made connections to computing.

*4.1. Teachers' Perceptions of Quantum*

We identified five themes related to teachers' perceptions of quantum: (a) teachers were generally excited and interested in quantum, (b) teachers saw the potential to advance equity in STEM through quantum teaching, (c) teachers identified systemic barriers to STEM, (d) teachers were concerned about whether their students were able to learn quantum content, and (e) teachers varied in their own sense of readiness to implement quantum activities in their classroom. Most teachers (46 of 49) expressed excitement about and interest in quantum with statements like, "I used to think quantum was a subject only understandable to adults, but I now think the information can be highly digestible and applicable to children". Other teachers noted, "Now I think I can do this!" and "It's going to be fun!" Teachers' survey responses were aligned with our own observations, as evidenced by field notes captured during and after the professional development workshops. One entry dated 28 January 2023 noted, "teachers are engaged and energetic. Worked on activities in mixed grade-level groups and discussed vertical articulation of content. Lots of laughter and eagerness to try different activities". We did not find any differences between elementary and secondary teachers related to this theme.

In a follow-up interview, one elementary teacher shared what she found most compelling about teaching quantum to elementary students. "You have to get their attention, and it has to be something kind of cool. There's too many other things they are bombarded with, so when we talk about science, it helps if it's a little bit flashy". She went on to explain one of her main takeaways from the professional development workshop, "Just not being scared of the word. Just like it's quantum, but there's all these concepts within it that we can explore, and there's a ton of things that we can do". This finding is important because it illustrates that K-12 teachers are excited about quantum and interested in finding ways to bring it into their classrooms.

Many of our teachers (17 of 49) made note of ways that introducing quantum may advance equity in STEM. We noticed some trends in this theme related to their grade level. Elementary teachers focused on how introducing quantum in their classrooms could increase equity and access. One elementary specialist noted, "Things are going to progress and develop rapidly in this field, and this could give our young students a leg up in this area of science". Another elementary teacher noted that teaching quantum "provides an opportunity for students to grapple with a different way of thinking". These teachers were excited about how learning about quantum offered new ways of learning for their students.

Some secondary teachers also noted how quantum could level the playing field for their diverse learners, including one who wrote, "English Learners have the same opportunities as native speakers to join STEM lessons, and they all start at the same knowledge level with quantum". A high school biology teacher who teaches English Learners was excited to share, "Being fluent in English is not a prerequisite for quantum as long as scaffolding and differentiation is available". Other teachers noted the novelty of quantum concepts as helping to include students who might not normally connect to their discipline. A secondary math teacher explained this, "It is totally untraditional. Now we are talking about electrons not having a specific location, but statistical probability of a location then comes in useful quadratic equations".

In our exit survey, we asked teachers to identify concerns about teaching quantum, and 40 of the 49 participating teachers did so. In our coding and classification of teachers' responses, we first focused on teachers' concerns about students that we classified as systemic. The most often expressed concern (n = 14) was related to teachers' uncertainty about a clear alignment to content standards. One sixth-grade teacher wrote, "Fitting quantum into standards is a concern. I would be concerned that it would only occur as an extension". Similarly, a fifth-grade teacher responded, "Having quantum accepted into the county's curriculum/pacing guide and having time to teach it". This teacher could only imagine teaching quantum if it was included in her school district's mandated curriculum.

Among the 14 teachers concerned with alignment to standards, five also expressed concern about time. These comments addressed having the time to teach quantum and/or to learn to teach quantum. A high school chemistry teacher shared, "We need TIME (and essentially money) to figure out how to incorporate the basic knowledge with quantum concepts. They SHOULD go together, but the Virginia Standards of Learning so far do not reflect them". A high school biology teacher also noted time constraints as a concern and then added, "I don't have a clear grasp of what my students should get out of me introducing quantum concepts to them. Are quantum concepts the vehicle, or are they the destination?". These teachers' responses illustrate the curricular constraints on teachers and how they serve as a systemic barrier to integrating quantum into K-12 classrooms.

Teachers (n = 11) also expressed concern about whether their students were able to learn quantum content. Several noted concerns about getting all of their students to learn, including an elementary teacher who shared that she was worried about "Reaching students at all academic levels in my general education classroom, especially those lower level students". Other elementary teachers were uncertain about their students' ability to understand the vocabulary used in quantum, including one teacher who stated she would have to "think about how to adapt and modify this content" for her English Learners. A middle school science teacher shared that she had already begun to think about addressing the potential language barrier, noting that she would "need to create visuals and more resources to explore these concepts in the English Learner classroom". Additional concerns about student ability related to student preparation for learning quantum. These concerns were expressed by two secondary teachers, who both believed their students lacked background knowledge in probability and probabilistic thinking.

Twelve of our forty-nine teachers (24%) expressed a lack of confidence in their readiness to implement a quantum activity within their classroom. We wondered if we would see a difference in teacher responses based on their grade level, but we had a proportional number of elementary and secondary teachers respond in this way. However, we found a notable difference based on the format of professional development attended by the teachers. The majority of teachers who participated in the five-hour in-person professional development (83%) expressed confidence in their ability to integrate quantum into their classrooms, including a middle school STEM teacher who noted, "It is inspiring how we can introduce these concepts to our students, start little by little exploring these concepts with our students and offer authentic learning experiences for our students". In contrast, only 54% of teachers who attended the two-hour online professional development felt ready to teach about quantum science and technologies in their classroom. One high school teacher who attended the online workshop and did not feel prepared to teach quantum explained, "I don't totally understand it myself and won't feel comfortable answering questions". A sixth-grade teacher wrote, "It's heavy stuff and a little over my head". Several teachers expressed skepticism that they would be able to find connections to the standards. They felt that they would not have the support of their administration to implement quantum without an explicit connection.

*4.2. Curricular Connections*

Approximately half of the participants (27 of 49) identified explicit connections to content within their standards. Sixteen of the teachers identified specific science content they would use to introduce or explore quantum concepts, including "rays and waves and atoms", the "atomic quantum model of the atom", "light waves", and the "physics of light". Ten teachers noted connections to mathematics content, with most connecting to probability, several connecting to measurement, and one to exponential growth. Four teachers identified computer science topics, including teaching binary for classical vs quantum computing and simply teaching about quantum computers and how they differ from classical computers. Only three teachers made connections from quantum to multiple disciplines.

Teachers also commented on the relevance and importance of teaching quantum. Several elementary teachers noted that, before the workshops, they thought quantum was not accessible for elementary students. Yet, they left the workshop convinced that they could try using the resources provided in the workshop. A secondary teacher also commented on this, writing, "I learned that quantum physics can be shared at any level. Having elementary teachers at the workshop wanting to share this material with their students made me want to share this with my high school students".

Several teachers in our study discussed the imperative of bringing quantum into K-12 classrooms. An elementary teacher stated this clearly, "we need to be introducing the basics to our students at a young age. Quantum can be for any age and will possibly play a major role in the lives of our youngest learners when they grow up".

## 5. Discussion and Conclusions

This study offers insight into teachers' perceptions of incorporating quantum concepts and technologies in K-12 classrooms. The teachers in our study expressed excitement about and interest in learning and teaching about quantum and believed that their students would also find it exciting. This finding about quantum is consistent with prior research indicating that teachers perceive STEM as inherently motivating [60] and offers encouragement to researchers, teacher educators, and policymakers who are considering whether K-12 teachers will embrace quantum ideas. It is likely that the nature of our professional development workshops influenced teachers' sense of excitement about quantum. We designed the workshops to build K-12 teachers' knowledge of intriguing quantum concepts and to engage teachers in hands-on activities that they could use with their own students. We positioned teachers to reflect on why quantum education would benefit their students and to discuss how it could fit within their curriculum.

Teachers in our study also saw quantum education as a novel method to advance equity in STEM. They found it appealing that their students would be unlikely to have previously encountered quantum concepts, as they would all be starting from the same place conceptually. Our participants' intuition is supported by prior research in science and computing that indicates emerging fields are often more open to under-represented and marginalized groups than once they are established [61,62].

The interview and survey responses from participants offered insight into teachers' perceptions of barriers to integrating quantum in K-12. These barriers included systemic barriers like time and standards alignment as well as teachers' perceptions of their students' preparation and ability to learn quantum and teachers' beliefs about their own readiness to teach quantum. Our findings are consistent with prior studies that identified teachers' perceptions of barriers to K-12 STEM integration [60,63,64]. This suggests that at least some teachers see quantum as similar to other STEM content and that quantum education researchers should be aware of the literature on K-12 STEM integration.

Some teachers in our current study expressed concerns about their readiness to teach quantum after their professional development experience, which supports the findings of Matsler and colleagues [34]. However, in our study, this was far more common in the teachers who attended only a brief online workshop. The majority of the teachers who attended the in-person extended workshop expressed confidence in their ability to incorporate quantum activities and concepts in their classrooms. It is clear that the length and format of the professional development plays a crucial role in teachers' sense of readiness.

Over half of our participants identified connections to their disciplinary content and curriculum. This must be considered in connection with our finding that nearly one-third of our participants were concerned with quantum's lack of alignment with their standards and curriculum. It is critical that teachers be able to imagine where quantum fits within their existing curriculum. The National Q-12 Education Partnership's [15] work to develop quantum education frameworks can help teachers understand how their discipline-specific learning goals can be met through teaching quantum concepts. Our previous work [65] with a small sample of teachers demonstrated that teachers draw on their experience and knowledge of the curriculum to find creative connections between quantum and the K-12 curriculum. Future work in this area should provide space for K-12 teachers to engage in dialogue with each other about quantum content, their standards, and how these may fit together.

Our findings suggest that it is possible to fit quantum concepts within existing state standards and curricula. However, it is also clear that the current systemic constraints on teachers due to overloaded curricula and the lack of time for "extra topics" will restrict the degree to which quantum concepts can be integrated into the K-12 curriculum. If we are to meet the future needs of our society and position students as knowledgeable and ready, it will be necessary to make quantum concepts explicit within K-12 STEM standards across the United States.

This study employed a critical realist methodological perspective and qualitative methods to investigate teachers' perceptions of quantum in K-12 schools. Qualitative research is inherently subjective, and we make no claims of generalizability. Critical realism assumes there are multiple understandings of reality [50]; our findings are specific to the professional development context in which data were collected and to the research team who conducted the research.

An additional limitation of our study is that the teacher participants chose to attend this quantum professional development workshop. We recruited participants through our partnerships with local school districts and accepted all teachers who expressed an interest in attending a workshop. Our participants may not represent typical K-12 teachers, so the findings of our study are not generalizable to all K-12 teachers in the U.S. In future research, we plan to partner with school districts to implement our quantum professional development with teachers with varying interests in quantum technologies and STEM.

This study contributes to the expanding body of research on what quantum can and should look like in K-12 schools. Although quantum content is considered by many to be particularly challenging and something that only "smart" people can understand [66], the teachers in this study approached quantum with an open mind. They had a vision for teaching quantum to students at all levels. Working with teachers to integrate quantum science and technologies into the K-12 curriculum is an opportunity to expand student access to meaningful integrated STEM and science content.

K-12 students must be prepared to utilize and contribute to the evolving quantum technologies that stand to transform computing, sensing, imaging, and communication. Teachers play a crucial role in providing opportunities for students to build the interest and essential knowledge necessary to participate in our quantum future.

**Author Contributions:** Conceptualization, N.H., J.L.R., T.N.B., and X.Z.; methodology, N.H. and J.L.R., data collection, M.C., J.L.R., and N.H.; formal analysis, N.H., T.N.B., and M.C.; writing—original draft preparation, N.H., T.N.B., and M.C.; writing—review and editing, N.H. and X.Z.; funding acquisition, J.L.R., N.H., and B.W.D. All authors have read and agreed to the published version of the manuscript.

**Funding:** This research was partially funded by the U.S. National Science Foundation (Award 2329874) and the U.S. Department of Education, Community Funded Projects (Award S215K220031). However, these contents do not necessarily represent the policy of the U.S. Department of Education, and you should not assume endorsement by the federal government.

**Institutional Review Board Statement:** The study was conducted in accordance with the Declaration of Helsinki and approved by the Institutional Review Board of George Mason University (Project 1811048, 23 September 2021).

**Informed Consent Statement:** Informed consent was obtained from all subjects involved in the study.

**Data Availability Statement:** The data presented in this study are available on request from the corresponding author. The data are not publicly available due to privacy concerns.

**Conflicts of Interest:** The authors declare no conflicts of interest. The funders had no role in the design of the study, in the collection, analyses, or interpretation of data, in the writing of the manuscript, or in the decision to publish the results.